\documentclass[aps,onecolumn]{revtex4}
\usepackage[colorlinks=true, pdfstartview=FitV, linkcolor=blue, citecolor=red, urlcolor=magenta, breaklinks=true]{hyperref}
\usepackage{color}
\usepackage{subeqnarray}
\usepackage{graphicx}
\usepackage[all]{xy}
\usepackage{amsmath}
\usepackage{amssymb}
\usepackage{subfigure}
\usepackage{mwe}
\newcommand{\be}{\begin{equation}}
\newcommand{\ee}{\end{equation}}
\newcommand{\ben}{\begin{eqnarray}}
\newcommand{\een}{\end{eqnarray}}
\newcommand{\bes}{\begin{subequations}}
\newcommand{\ees}{\end{subequations}}
\newcommand{\bens}{\begin{subeqnarray}}
\newcommand{\eens}{\end{subeqnarray}}

\newcommand{\bb}{\bibitem}
\def\tanh{\text{tanh}}
\def\sech{\text{sech}}
\def\sn{\text{sn}}
\def\dn{\text{dn}}
\def\cn{\text{cn}}
\def\sgn{\text{sgn}}

\begin{document}
\title{New results on asymmetric thick branes}
\author{D. Bazeia}\email{dbazeia@gmail.com}
\affiliation{Departamento de F\'\i sica, Universidade Federal da Para\'\i ba, 58051-970 Jo\~ao Pessoa, PB, Brazil}

\author{D. A. Ferreira}\email{ferreira.douglasdaf@gmail.com }
\affiliation{Departamento de F\'\i sica, Universidade Federal da Para\'\i ba, 58051-970 Jo\~ao Pessoa, PB, Brazil}

\vspace{3cm}
\date{\today}

\begin{abstract}

This work deals with the presence and stability of thick brane solutions in the warped five dimensional braneworld scenario with a single extra spatial dimension of infinite extent. We combine two distinct procedures that give rise to new possibilities, allowing that we describe models of asymmetric thick branes, with the asymmetry being controlled by a single real parameter. We illustrate the main results with some distinct models, which show that the method works for both standard and generalized models, and the solutions are gravitationally stable against small perturbations of the metric.

\end{abstract}


\maketitle

\section{Introduction}

The braneworld scenario in which one deals with a warped geometry that engenders a single extra spatial dimension of infinite extent was initially proposed in Ref.~{\cite{Ra}}. In this case, the brane is considered to be geometrically thin (thin brane) and embedded in five dimensional spacetime.
However, since the above scenario is motivated by a unification theory, there seems to be a minimum length scale and it is also of interest to consider the thickness of the brane. For this reason, thick brane scenarios were considered in~{\cite{GW,De,Csa,Gre}} where the warp factor behaves smoothly due to the presence of a scalar field source in the Einstein-Hilbert action with the warped geometry. Since then, thick brane models have been widely investigated by several authours with distinct motivation, see for example~{\cite{Cam,Ko,Ba,Ba2,Bar,Du,Zhong,Pey}} and ~{\cite{Fo}} for a review.

An interesting issue about the thick brane scenario is that it may have an asymmetric profile, when the scalar potential has no symmetry $ Z_{2} $. In this situation, the brane can asymptotically connect different five-dimensional spacetimes. The asymmetric feature of the brane have been studied in various contexts~{\cite{Pa,Pa2,Me,Du2,Ba3,Ahmed,Ba4,Mene}}. For example, in Refs.~{\cite{Pa,Pa2}} the authors have used asymmetric brane models to explain the late-time acceleration of the universe from infrared modifications of gravity. In addition, the Friedmann equations were obtained for an $ AdS_{5} $ bulk, with different cosmological constants on the two sides of the brane in~{\cite{Me}}. More recently, an interesting model so-called asymmetric bloch brane has been offered, under appropriate assumption, to address the hierarchy problem~{\cite{Du2}}. Other works have dealt with the construction of new models for asymmetric braneworld scenario~{\cite{Ba3,Ahmed,Ba4,Mene}} and this has motivated us to implement the current study.

In this sense the aim of this work is to construct new models for asymmetric thick branes using the procedure described in \cite{Ba3} and in \cite{Ahmed} in addition to the method developed in Ref.~{\cite{Du3}}, which consists in considering that the effective superpotential is a sum of $ n $ decoupled superpotentials, with $n$ accounting for the number of independent scalar fields that appear in the model. In Minkowski spacetime this case represents a trivial situation but nontrivial results arise when we deal with braneworld scenarios, since in curved spacetime there appear another contribution to the scalar potential that couples the scalar fields and introduces extra effects. 

The investigation to be implemented in this work is organized as follows. In Sec.~\ref{sec2} we briefly review the main results on thick branes such as the equations for the warp function and the energy density. In Sec.~\ref {sec3}, we introduce the procedure and illustrate it with several examples. We then examine the stability of the models in Sec.~\ref{sec4} and show in Sec.~\ref{sec5} that the procedure developed here can also be implemented in models with generalized dynamics. Finally, in Sec.~\ref{sec6}, we present our concluding remarks.

\section{Braneworld Scenario} \label{sec2}
Let us now investigate scalar fields in the braneworld scenario with a single spatial dimension of infinite extent, described by the line element
\be \label{eq3.1}
ds^{2}_{5}=e^{2A}\eta_{\mu\nu}dx^{\mu}dx^{\nu}-dy^{2}\,,
\ee
where $ A=A(y) $ is the warp function, $ y $ is the extra dimension and $\eta_{\mu\nu}$ describes the four-dimensional Minkowski spacetime $(\mu,\nu=0,1,2,3)$. In this case, the action is given by
\be\nonumber
I=\int dx^{4}dy\sqrt{\vert g\vert}\left(-\frac{R}{4}+{\cal L}\right);
\ee
\be \label{eq3.2}
{\cal L}=\frac{1}{2}\sum_{i=1}^{n}g_{ab}\partial^{a}\phi_{i}\partial^{b}\phi_{i}-V(\phi_{1},...,\phi_{n})\,,
\ee
with $a,b=0,1,2,3,4$ and $ \phi_{i}, i=1,2,...,n$ stands for the scalar fields. Also, we are using natural units and $4\pi G_{5}=1$, for simplicity. The minimization of the previous action with respect to the metric allows us to write Einstein's equations
\be \label{eq3.3}
G_{ab}=2\,T_{ab}\,.
\ee
Considering that the scalar fields and the warp function depend only on the extra dimension, we get
\be \label{eq3.4}
A''=-\frac{2}{3}\sum_{i=1}^{n}\phi_{i}'^{2}\,,
\ee
\be \label{eq3.5}
A'^{2}=\frac{1}{6}\sum_{i=1}^{n}\phi_{i}'^{2}-\frac{1}{3}V(\phi_{1},...,\phi_{n})\,.
\ee
Here the prime represents derivative with respect to $y$; that is $A^\prime=dA/dy$, etc. We can also implement a first order formalism considering that the potential has the form
\be \label{eq3.6}
V(\phi_{1},...,\phi_{n})=\frac{1}{2}\sum_{i=1}^{n}W_{\phi_{i}}^{2}-\frac{4}{3}W^{2}\,,
\ee
with $W=W(\phi_{1},...,\phi_{n})$ being the superpotential and $W_{\phi_i}=dW/d\phi_i$. Consequently, we obtain the following first order differential equations
\be \label{eq3.7}
\phi_{i}'=W_{\phi_{i}}\,,
\ee
\be \label{eq3.8}
A'=-\frac{2}{3}W(\phi_{i})\,,
\ee
which solve the equations \eqref{eq3.4} and \eqref{eq3.5}. 

Furthermore, we can show that the energy density which is given by
\be \label{eq3.9}
\rho(y)=e^{2A}\left(\frac{1}{2}\sum_{i=1}^{n}\phi_{i}'^{2}+V(\phi_{1},...,\phi_{n})\right)\,,
\ee
can be rewritten in the form
\be \label{eq3.10}
\rho(y)=\dfrac{d}{dy}(e^{2A}W)\,.
\ee
The energy of this system is found by performing the integration with respect to the extra dimension, so we get that the energy of the brane associated with the scalar fields is null.

\section{The procedure}\label{sec3}
In this section we combine the approach implemented in \cite{Ba3} and \cite{Ahmed} with the method proposed in Ref.~{\cite{Du3}} to propose and investigate new braneworld models. We then write the superpotential in the form
\ben\label{w}
W(\phi_{1},...\phi_{n})&=&k+ \sum_{i=1}^{n}W_{i}(\phi_i)\nonumber\\
&=&k+W_1(\phi_1)+\cdots+W_n(\phi_n)\,,
\een
where $k$ is a real constant. In this sense, although $\partial W/\partial\phi_i $ only depends on $ \phi_{i} $, the potential will couple the several scalar fields due to the second term in Eq. \eqref{eq3.6}. Moreover, an interesting feature of this superpotential is that it keeps the first order equations \eqref{eq3.7} decoupled. An immediate consequence of the above procedure is that it generalizes the results presented in Ref.~{\cite{Ba3}}, where the models were described by a single scalar field.

Although the procedure described above is robust enough to describe $ n $ scalar fields, let us investigate, for simplicity, the case $ n=2 $ with $ \phi_{1}=\phi $ e $ \phi_{2}=\chi  $. We illustrate this possibility with several models that follow below.
   
\subsection{Model 1}
We consider a model of two scalar fields described by the superpotential
\be \label{eq4.1}
W(\phi,\chi)=k+ \left(\phi-\frac{\phi^{3}}{3}\right)+\left(\chi-\frac{\chi^{3}}{3}\right)\,,
\ee
which combines two models of the Higgs type. In this case we get the potential
\be
V(\phi,\chi)=\frac{1}{2}\big((1-\phi^{2})^{2}+(1-\chi^{2})^2\big)-\frac{4}{3}\bigg(\phi-\frac{\phi^{3}}{3}+\chi-\frac{\chi^{3}}{3}+k \bigg)^{2}\,,
\ee
and the first order equations \eqref{eq3.7} become
\be \label{eq4.2}
\dfrac{d\phi}{dy}=(1-\phi^{2})\,,\,\,\,\,\,\dfrac{d\chi}{dy}=(1-\chi^{2})\,.
\ee
\begin{figure}[t]
\centering
\begin{subfigure}
\centering
{\includegraphics[width=6.5cm,height=4.5cm]{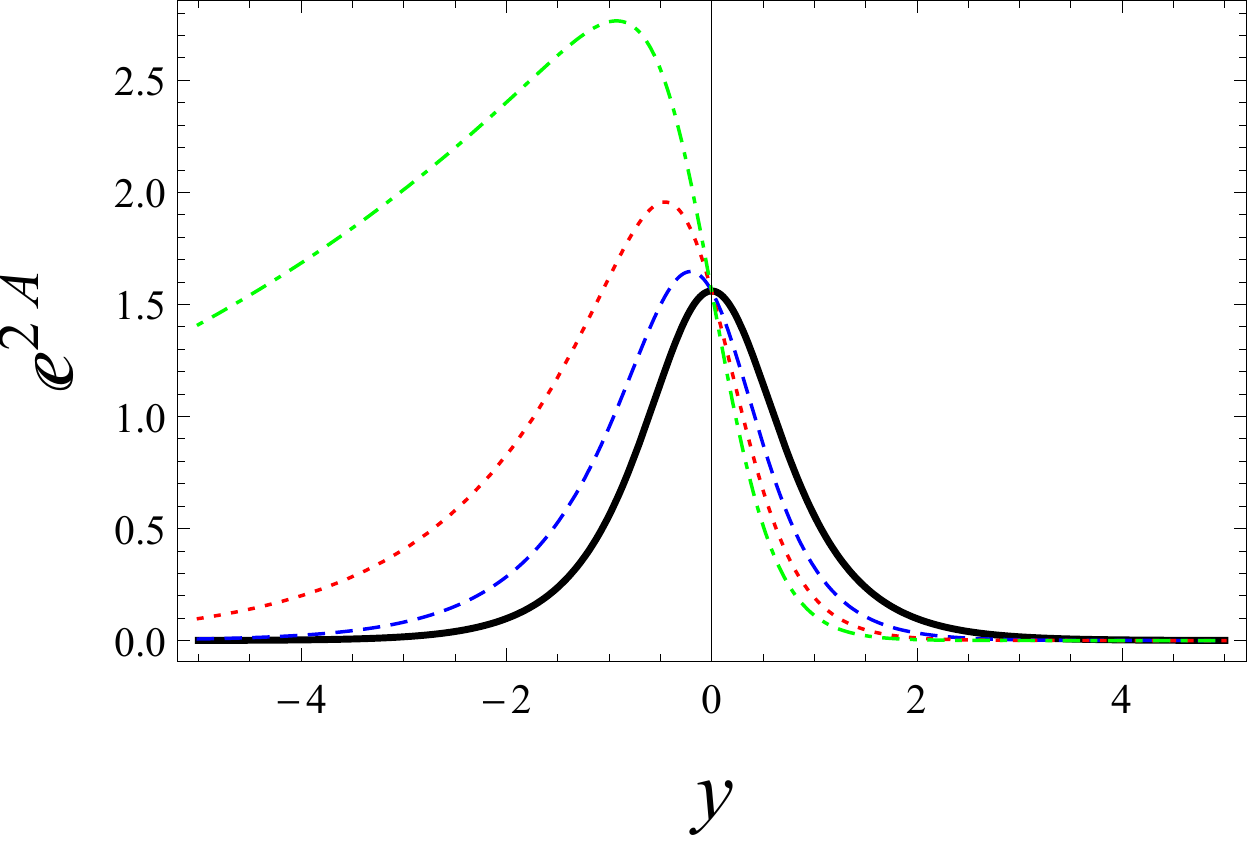}}
\end{subfigure}
\qquad
\begin{subfigure}
{\includegraphics[width=6.5cm,height=4.5cm]{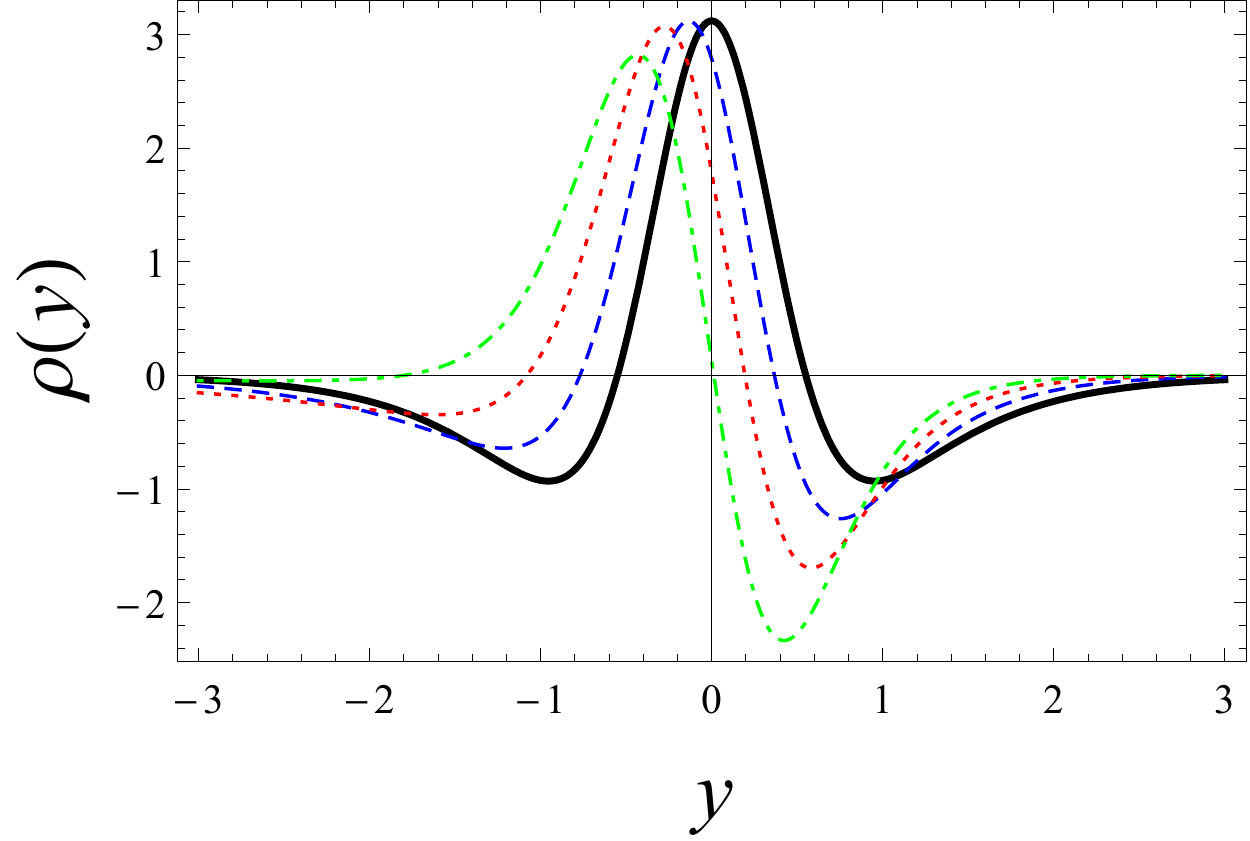}}
\end{subfigure}
\caption{The model 1. Warp factor (left panel) and  energy density (right panel) for $ k=0 $ (black, solid line), $ k=0.4 $ (blue, dashed line), $ k=0.8 $ (red, dotted line) and $ k=1.2 $ (green, dot-dashed line).}\label{Fig1}
\end{figure}

The solutions of the above equations are given by
\be \label{eq4.3}
\phi(y)=\tanh(y)\,,\,\,\,\,\,\,\,\chi(y)=\tanh(y)\,,
\ee
where the minima are $ \phi_{\pm}=\chi_{\pm}=\pm1 $. In this case the equation \eqref{eq3.8} becomes
\be \label{eq4.4}
\dfrac{dA}{dy}=-\frac{2}{9}\big(3k+6\tanh(y)-2\tanh^{3}(y)\big)\,.
\ee
By integrating the above equation we get the following warp function
\be \label{eq4.5}
A(y)=-\frac{2}{9}(3ky+4\ln(\cosh(y))-\sech^{2}(y))\,.
\ee
Note that the behavior of this function for very large values of $y$ can be written as
\be \label{eq4.6}
A_{asym\pm}(y)=-\frac{2}{3}W(\phi_{\pm},\chi_{\pm})|y|=-\frac{2}{3}\left(k\pm\frac{4}{3}\right)|y|\,.
\ee
Furthermore, we find that the asymptotic behavior in the limits $ y\rightarrow\pm \infty $, defines the five-dimensional cosmological constant 
\be \label{eq4.7}
V(\phi_{\pm},\chi_{\pm})\equiv \Lambda_{5\pm}=-\frac{4}{3}\left(k\pm\frac{4}{3}\right)^{2}\,.
\ee
For $ k=0 $ (symmetric case) we have that both sides of the brane are connected by the same cosmological constant. When $ 0<\vert k \vert<4/3 $ the model has different cosmological constants on each side of the brane connecting different $ AdS_{5} $ bulk spaces. For other values of $ k $ it diverges, leading to no physically acceptable braneworld scenario.

However, it is worth mentioning that an interesting issue arises when $ \vert k \vert=4/3 $. In this case on one side of the brane there is a negative cosmological constant and as a result the bulk should be $ AdS_{5} $. But on the other side of the brane the cosmological constant of bulk vanish, so the bulk is asymptotically Minkoskwi $ (\mathbb{M}_{5})$. We remember here that situations similar to this, where space is asymptotically 5D AdS-Minkoskwi, appears very naturally in the context of supergravity~{\cite{Cve}}.

Using Eq.\eqref{eq4.5} we can write the warp factor explicitly in the form
\be \label{eq4.8}
e^{2A(y)}=e^{-\frac{4}{9}(3ky+4\ln(\cosh(y))-\sech^{2}(y))}\,.
\ee
It is depicted in the top panel of Fig. \ref{Fig1} for some values of $ k $, with the solid curve representing the case $ k = 0 $ (symmetric profile). The other curves represent $ k $ in the region where the brane connects a bulk $ AdS_{5} $ with different cosmological constants.
\begin{figure}[t]
\centering
\begin{subfigure}
\centering
{\includegraphics[width=6.5cm,height=4.5cm]{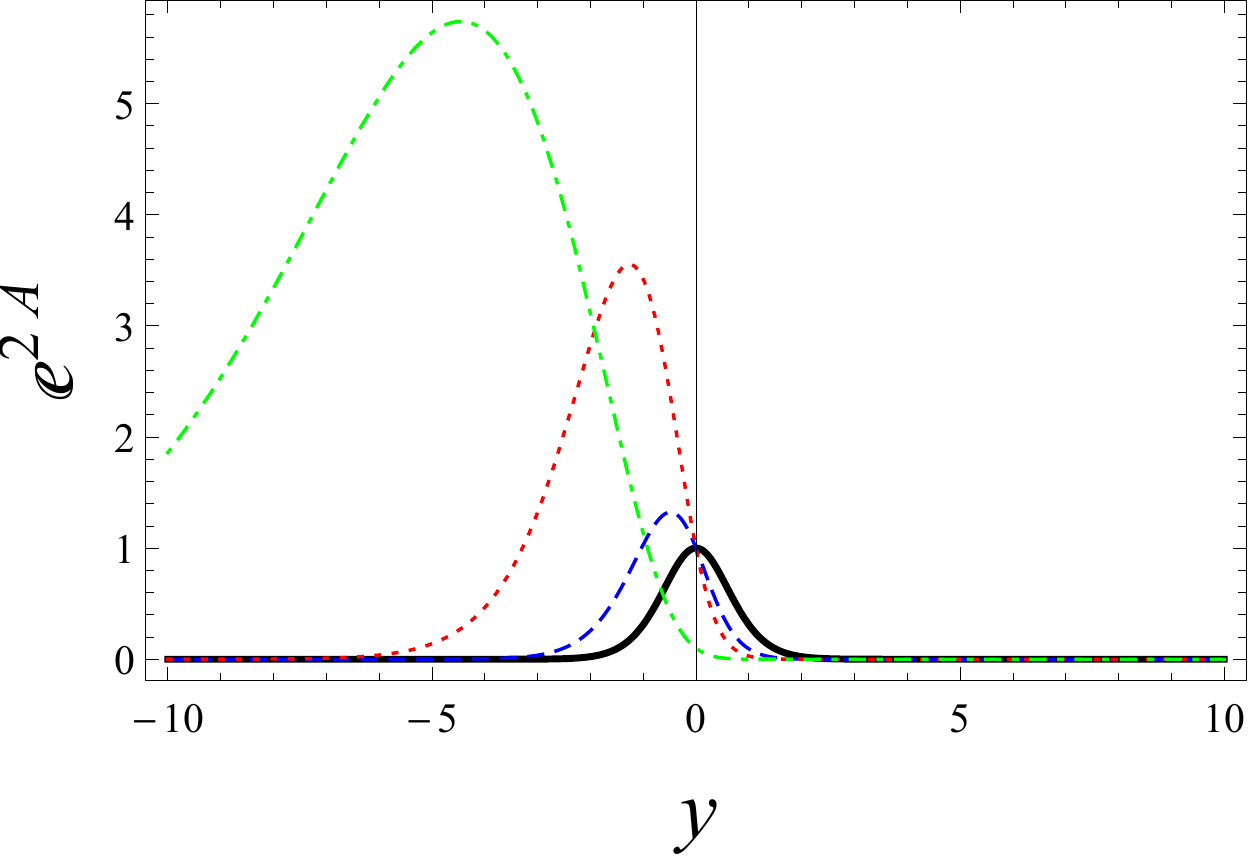}}
\end{subfigure}
\qquad
\begin{subfigure}
{\includegraphics[width=6.5cm,height=4.5cm]{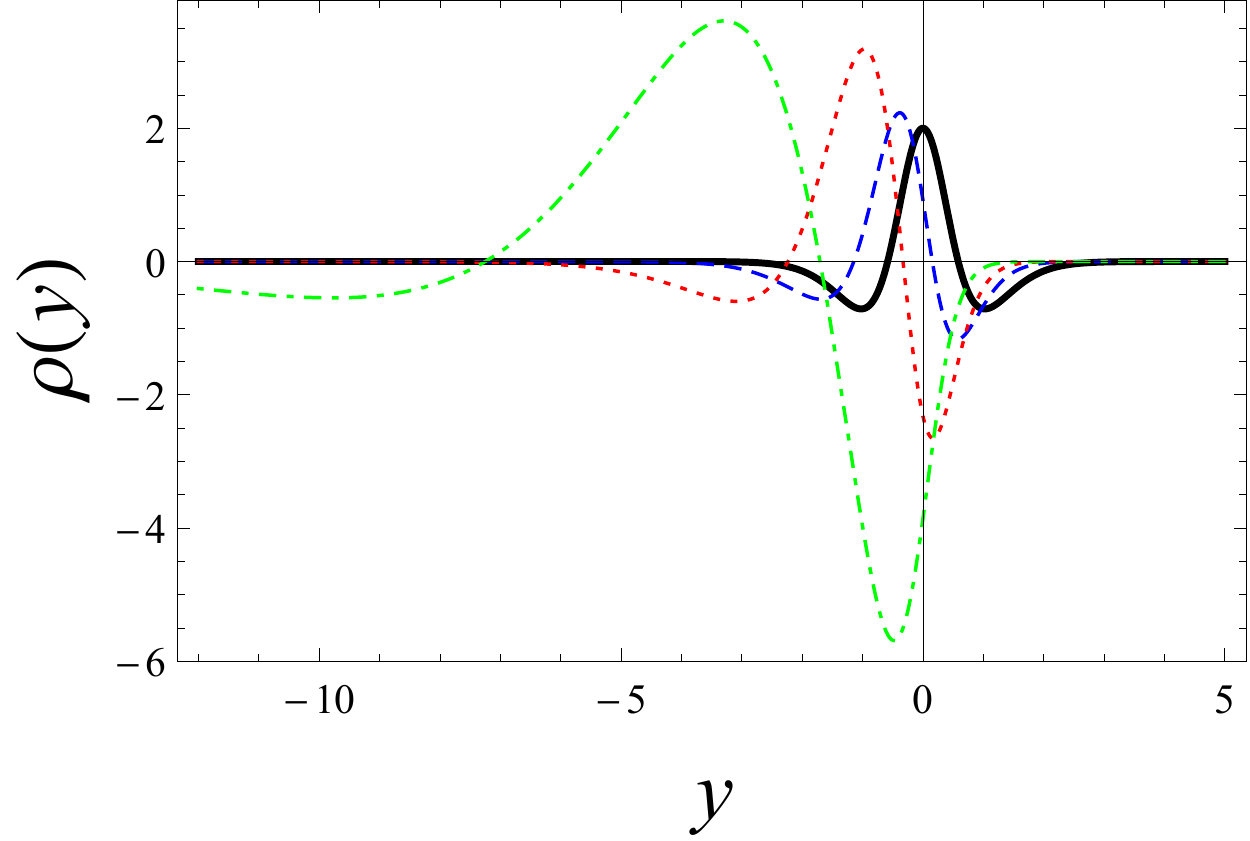}}
\end{subfigure}
\caption{The model 2. Warp factor (left panel) and  energy density (right panel) for $ k=0 $ (black, solid line), $ k=0.9 $ (blue, dashed line), $ k=1.8 $ (red, dotted line) and $ k=2.7 $ (green, dot-dashed line). For the latter case we depicted $ \exp(2A(y))/10 $ and $ \rho(y)/2 $. }\label{Fig2}
\end{figure}
On the other hand, the energy density can be obtained from Eq. \eqref{eq3.10} but as the complete expression is awkward we omit it here. However, in the bottom panel of Fig. \eqref{Fig1} we depict it for same values of $k$. Evidently, we see that the addition of a constant to the superpotential plays a crucial role in the construction of new models for asymmetric thick branes.

\subsection{Model 2}
Let us now deal with a model inspired by vacuumless system, where topological structures can appear in potentials without vacuum state. This was studied before in {\cite{Vi,Ba5}}, and here the model is described by
\be \label{eq5.1}
W(\phi,\chi)=k+ \tan^{-1}\left(\sinh(\phi)\right)+\tan^{-1}\left(\sinh(\chi)\right)\,.
\ee
Hence, the Eq. \eqref{eq3.6} leads to the following potential
\be \label{eq5.1.a}
V(\phi,\chi)=\frac{1}{2}\left(\sech^{2}(\phi)+\sech^{2}(\chi)\right)-\frac{4}{3}\left(\tan^{-1}(\sinh(\phi))+\tan^{-1}(\sinh(\chi))+k\right)^{2}\,. 
\ee
By substituting Eq. \eqref{eq5.1} in Eqs. \eqref{eq3.7} and solving the integrals, we obtain the solutions
\be \label{eq5.2}
\phi(y)=\sinh^{-1}(y)\,,\,\,\,\,\,\,\,\chi(y)=\sinh^{-1}(y)\,,
\ee
where the minima are given by $ \phi_{\pm}=\chi_{\pm}=\pm\infty $. Moreover, we see that the behavior of the superpotential in the minima results in $ W(\phi_{\pm},\chi_{\pm})=\left(k\pm\pi\right) $.

On the other hand the equation \eqref{eq3.8} becomes
\be \label{eq5.3}
\dfrac{dA}{dy}=-\frac{2}{3}\left(k+2\tan^{-1}(y)\right)\,.
\ee
Consequently, the integration of the above equation leads us to 
\be \label{eq5.4}
A(y)=-\frac{2}{3}\left(ky+2y\tan^{-1}(y)-\ln(1+y^{2})\right)\,,
\ee
where $ A(0)=0 $. In this case, the behavior of this function for very large values of $y$ is given by
\be
A_{asym\pm}(y)=-\frac{2}{3}\left(k\pm\pi\right)|y|\,,
\ee
and the five-dimensional cosmological constant is
\be
\Lambda_{5\pm}=-\frac{4}{3}\left(k\pm\pi\right)^{2}\,.
\ee
For $ k=0 $ (symmetric case) we have that both sides of the brane are connected by the same cosmological constant. For $ \vert k \vert=\pi $ on one side of the brane there is a negative cosmological constant and as a result the bulk should be $ AdS_{5} $ while the other side of the brane the bulk is asymptotically Minkoskwi $ (\mathbb{M}_{5})$. In the case where $ 0<\vert k \vert<\pi $ the model has different cosmological constants on each side of the brane connecting different $ AdS_{5} $ bulk spaces. For other values of $ k $ it diverges, leading to no physically acceptable braneworld scenario.

\begin{figure}[t]
\centering
\begin{subfigure}
\centering
{\includegraphics[width=6.5cm,height=4.5cm]{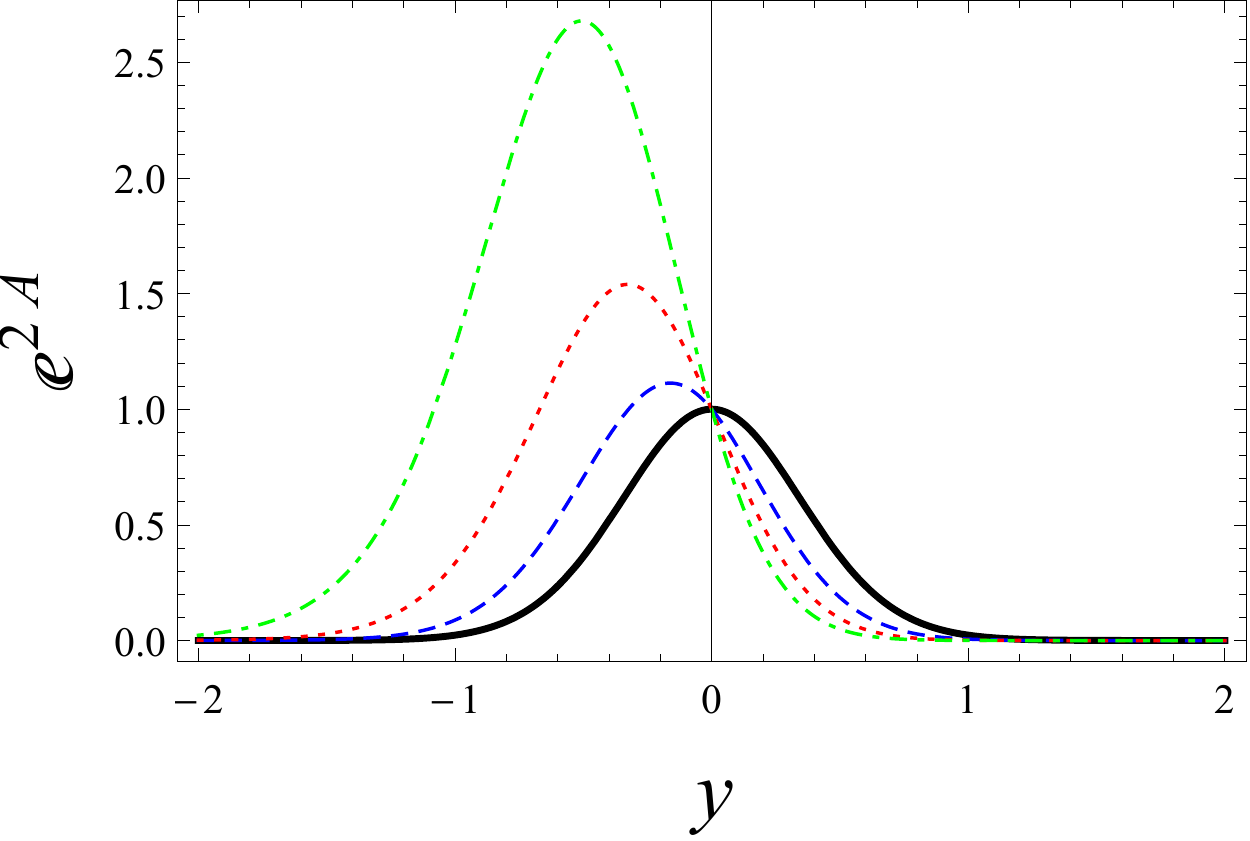}}
\end{subfigure}
\qquad
\begin{subfigure}
{\includegraphics[width=6.5cm,height=4.5cm]{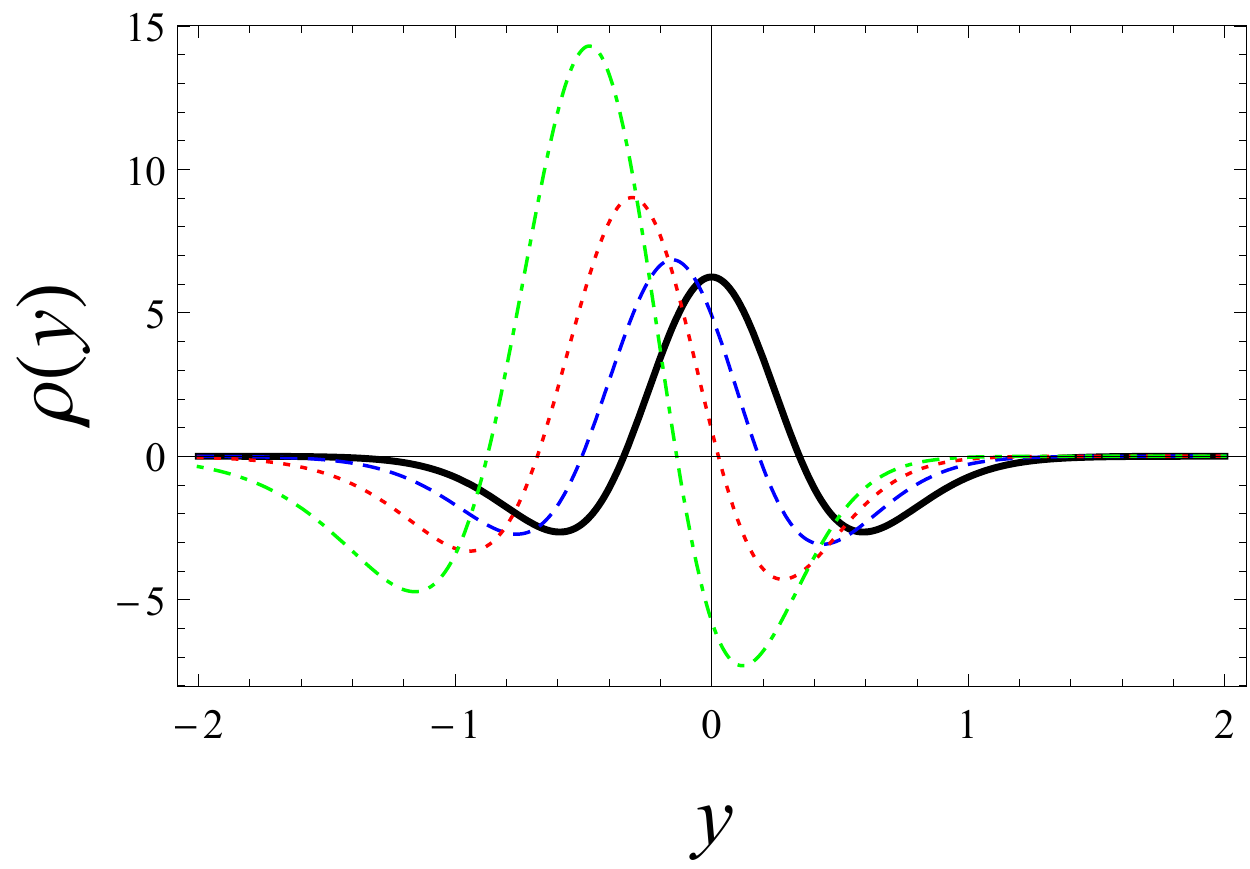}}
\end{subfigure}
\caption{The model 3 for $\lambda=0.6$. Warp factor (left panel) and  energy density (right panel) for $ k=0 $ (black, solid line), $ k=1 $ (blue, dashed line), $ k=2 $ (red, dotted line) and $ k=3 $ (green, dot-dashed line).}\label{Fig3}
\end{figure}

From Eq. \eqref{eq5.4}, we find the warp factor that is depicted in the top panel Fig. \ref{Fig2}, from which we can see the asymmetric behavior of the thick brane. By combining Eq. \eqref{eq3.10} with Eqs. \eqref{eq5.1} and \eqref{eq5.4}, we get the energy density of this model, whose asymmetric profile may be seen in the bottom panel of Fig. \ref{Fig2}. In both cases we use several values of $ k $ in the region where the brane connects a bulk $ AdS_{5} $ with different cosmological constants. 

\subsection{Model 3}
Let us now deal with a model inspired by Ref.~{\cite{More}}. The superpotential is given by
\be \label{eq6.1}
W(\phi,\chi)=k-\frac{1}{(1-\lambda_{1}) \sqrt{\lambda_{1}}} \ln \bigg(\frac{1-\sqrt{\lambda_{1}}\,\sn(\phi,\lambda_{1})}{\dn(\phi,\lambda_{1})}\bigg)-\frac{1}{(1-\lambda_{2}) \sqrt{\lambda_{2}}} \ln \bigg(\frac{1-\sqrt{\lambda_{2}}\,\sn(\chi,\lambda_{2})}{\dn(\chi,\lambda_{2})}\bigg)\,,
\ee
where $ \sn(\phi_{i},\lambda_{i}) $, $ \dn(\phi_{i},\lambda_{i}) $ are Jacobi's elliptic functions. We remember here that there is a set of basic elliptic functions denoted by $ \sn(\phi_{i},\lambda_{i}) $, $ \dn(\phi_{i},\lambda_{i}) $, $ \cn(\phi_{i},\lambda_{i}) $ and other, that obey interesting properties; see, e.g., Ref. \cite{GR} for more on this issue. Also, we are considering that both $ \lambda_{1} $ and $\lambda_2$  vary  in the interval $\in  [0,1) $.

From Eq. \eqref{eq6.1} we obtain the following potential
\ben
V(\phi,\chi)\!\!&=&\!\!\frac{1}{2}\bigg(\frac{1}{1-\lambda_{1}}\frac{\cn(\phi,\lambda_{1})}{\dn(\phi,\lambda_{1})}\bigg)^{\!2}\!\!\!+\!\frac{1}{2}\bigg(\frac{1}{1-\lambda_{2}}\frac{\cn(\chi,\lambda_{2})}{\dn(\chi,\lambda_{2})}\bigg)^{\!2}\;\;\;\nonumber \\
&&\!\!-\frac{4}{3}\bigg(k-\frac{1}{(1-\lambda_{1}) \sqrt{\lambda_{1}}} \ln \bigg(\frac{1-\sqrt{\lambda_{1}}\,\sn(\phi,\lambda_{1})}{\dn(\phi,\lambda_{1})}\bigg)-\frac{1}{(1-\lambda_{2}) \sqrt{\lambda_{2}}} \ln \bigg(\frac{1-\sqrt{\lambda_{2}}\,\sn(\chi,\lambda_{2})}{\dn(\chi,\lambda_{2})}\bigg)\bigg)^{\!\!2}\!\!\,.
\een 
It is straightforward to obtain the solutions using the first order equation \eqref{eq3.7}; they are
\be \label{eq6.21}
\phi(y)=\sn^{-1}\left(\tanh\big(y/(1-\lambda_{1})\big),\lambda_{1}\right)\,,
\ee
\be \label{eq6.22}
\chi(y)=\sn^{-1}\left(\tanh\big(y/(1-\lambda_{2})\big),\lambda_{2}\right)\,.
\ee
 
\begin{figure}[t]
\centering
\begin{subfigure}
\centering
{\includegraphics[width=6.5cm,height=4.5cm]{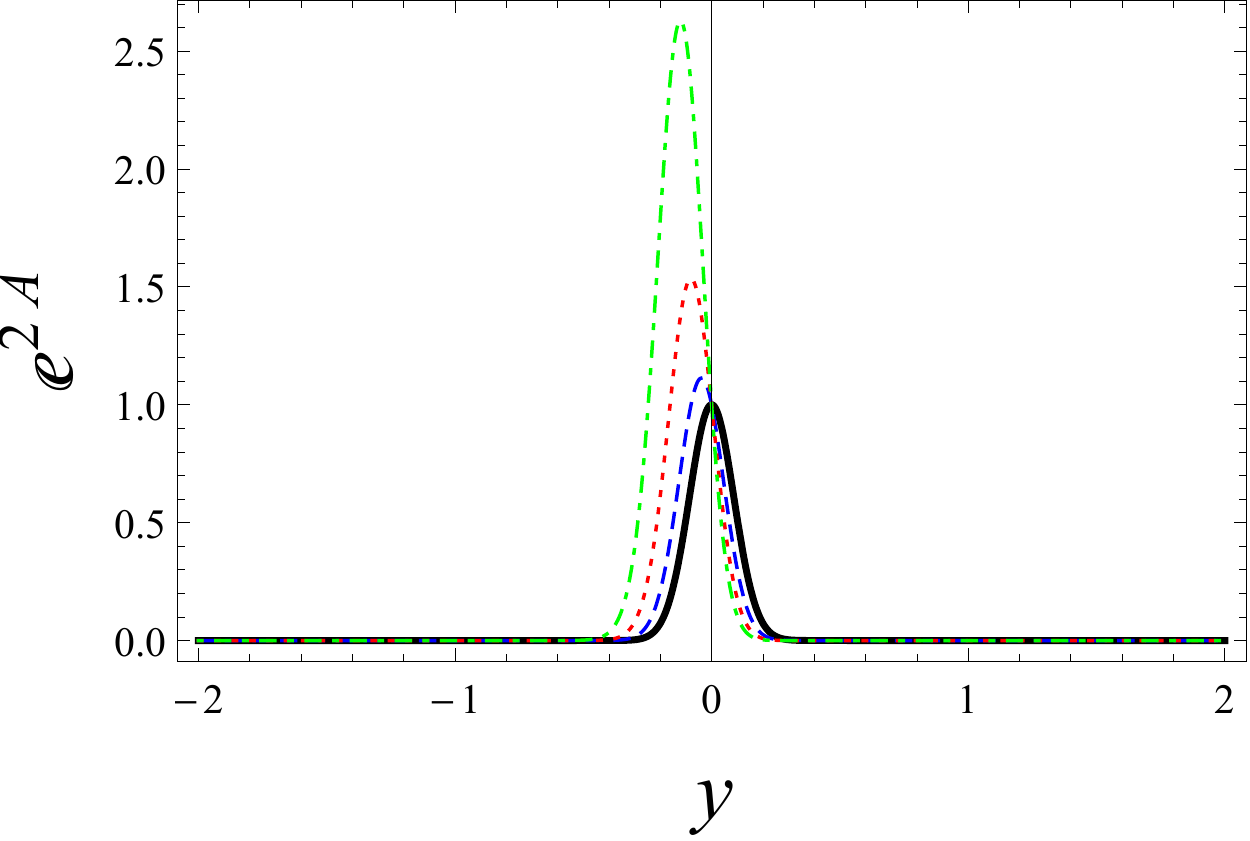}}
\end{subfigure}
\qquad
\begin{subfigure}
{\includegraphics[width=6.5cm,height=4.5cm]{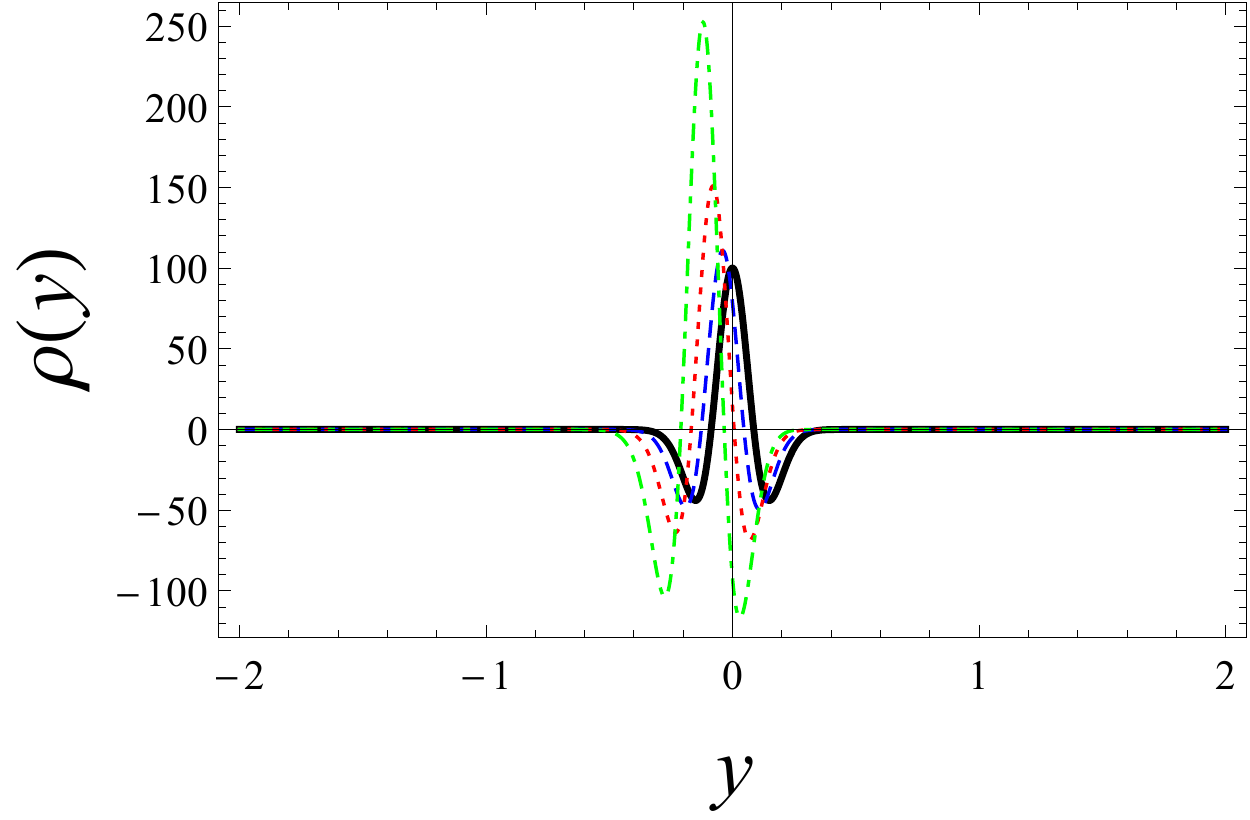}}
\end{subfigure}
\caption{The model 3 for $\lambda=0.9$. Warp factor (left panel) and  energy density (right panel) for $ k=0 $ (black, solid line), $ k=4 $ (blue, dashed line), $ k=8 $ (red, dotted line) and $ k=12 $ (green, dot-dashed line).}\label{figX}
\end{figure}

\begin{figure}[t!]
\includegraphics[width=7cm]{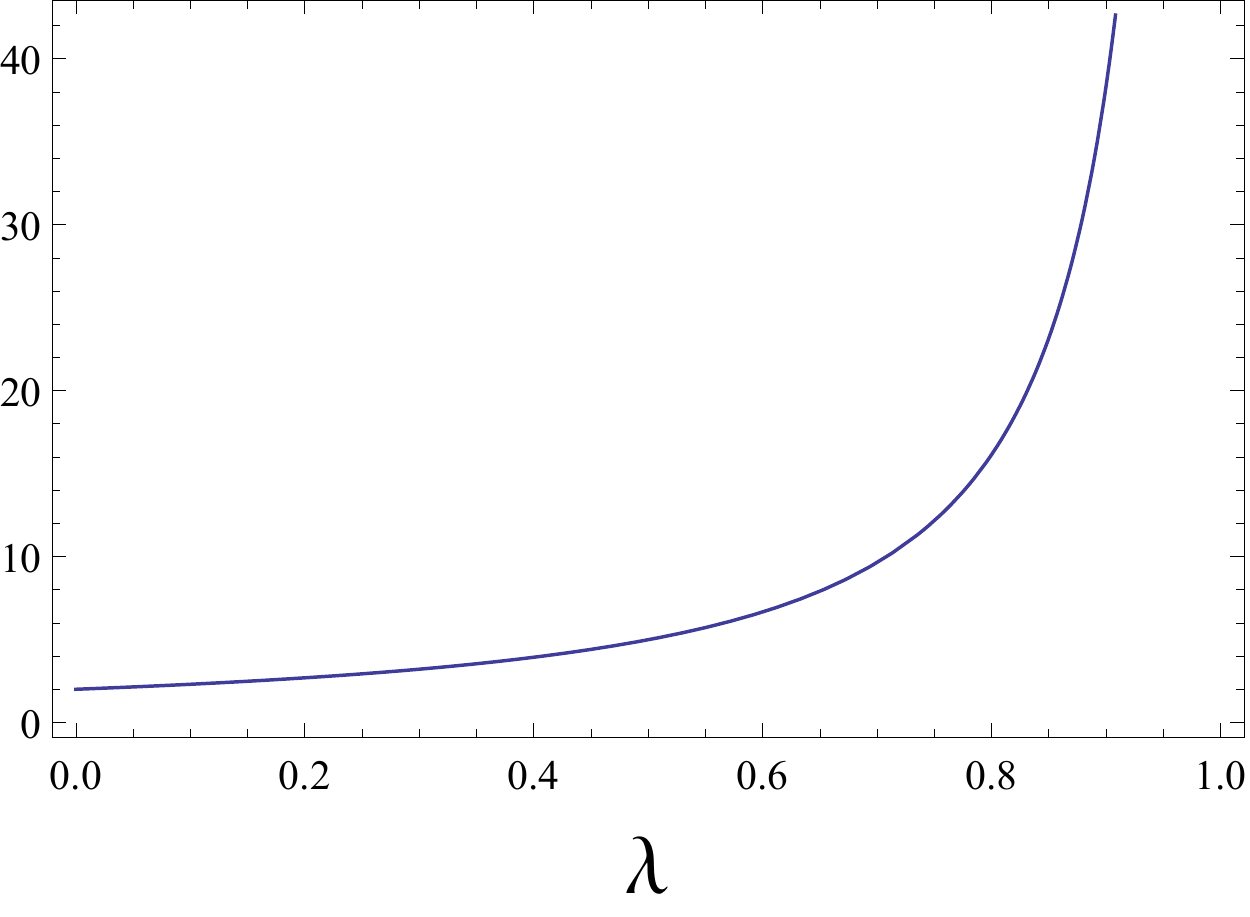}
\caption{The parameter $ k $ as a function of $ \lambda $, as it appears in Eq. \eqref{kl}.}\label{Fig4}
\end{figure}
\noindent Here, the minima are $ \phi_{\pm}=\pm K(\lambda_{1}) $ and $ \chi_{\pm}=\pm K(\lambda_{2}) $ with  $ K(\lambda_{1}) $ and $ K(\lambda_{2}) $ representing the complete elliptic integral of the first kind.

We take for simplicity the case with $ \lambda_{1}=\lambda_{2}=\lambda $, and now the equation \eqref{eq3.8} becomes
\be \label{eq6.3}
\dfrac{dA}{dy}=-\frac{2}{3}\bigg(k-\frac{2}{(1-\lambda)\sqrt{\lambda}}\ln \bigg(\frac{\dn\big(\sn^{-1}\big(\tanh(y/(1-\lambda)),\lambda\big),\lambda\big)}{1+\sqrt{\lambda}\,\tanh(y/(1-\lambda))}\bigg)\bigg)\,.
\ee
We have been unable to find the warp function analytically, so we get it numerically using the condition
$A(0)=0 $. 

We depict the results in Fig. \ref{Fig3} for $\lambda =0.6$ and in Fig. \ref{figX} for $\lambda =0.9$, for several values of $k$. The warp factor is in the top panels in Figs. \ref{Fig3} and \ref{figX}, from which we can see the asymmetric behavior of the thick brane which depends on $k$ and the shrinking effect that appears as we increase $\lambda$. The energy density of this model also exhibits an asymmetric profile which depends on $k$ and an shrinking effect which depends on $\lambda$, with its maximum peak increasing as
$k$ increases. These properties are shown in the bottom panels of Figs. \ref{Fig3} and \ref{figX}. In both cases we use different values of
$k$ in the region where the brane describes a bulk $AdS_{5}$ with different cosmological constants. We depicted the two
Figs. \ref{Fig3} and \ref{figX} at the same $y$ scale, to highlight the asymmetry behavior due to $k$ and the shrinking effect due to $\lambda$.

However, we can still see the analytical behavior of the warp function for very larger values of $y$; it has the form
\be
A_{asym\pm}(y)=-\frac{2}{3}\left(k\pm\frac{2}{(\lambda-1)\sqrt{\lambda}}\ln\left(\frac{\sqrt{1-\lambda}}{1+\sqrt{\lambda}}\right)\right)|y|\,,
\ee
and the five-dimensional cosmological constant is given by
\be
\Lambda_{5\pm}=-\frac{4}{3}\left(k\pm\frac{2}{(\lambda-1)\sqrt{\lambda}}\ln\left(\frac{\sqrt{1-\lambda}}{1+\sqrt{\lambda}}\right)\right)^{2}\,.
\ee
In this situation we see that the constant $ k $ depends on the value of $ \lambda $, and we get
\be\label{kl}
k=\frac{2}{(\lambda-1)\sqrt{\lambda}}\ln\left(\frac{\sqrt{1-\lambda}}{1+\sqrt{\lambda}}\right),
\ee
to give a vanishing cosmological constant in one of the two distinct sides of the brane. This is depicted in Fig. \ref{Fig4}. To illustrate, let us consider the case $ \lambda=0.6 $. Thus we have for $ k=0 $ (symmetric case) that both sides of the brane are connected by the same cosmological constant. For $ \vert k \vert\simeq6.66 $ on one side of the brane there is a negative cosmological constant and as a result the bulk should be $ AdS_{5} $ while the other side of the brane the bulk is asymptotically Minkoskwi $ (\mathbb{M}_{5})$. For $ 0<\vert k \vert<6.66 $ the model has different cosmological constants on each side of the brane connecting different $ AdS_{5} $ bulk spaces. For values outside this region the warp factor diverges, also leading to no physically acceptable braneworld scenario. For
$\lambda=0.9$ we get other values, but the calculations are similar. 

We can also consider other values of $\lambda$. As it was shown in \cite{More}, the interesting behavior is the shrinking of the braneworld solution as $\lambda$ increases toward unity. To see how this effect appears in the above model, we considered the cases of $\lambda=0.6$ and $\lambda=0.9$. 

\section{Stability}\label{sec4}
In order to study the stability of the gravity sector we must consider small perturbations in the metric in the form
\be \label{eq31.1}
ds^{2}=e^{2A}(\eta_{\mu\nu}+\epsilon h_{\mu\nu})dx^{\mu}dx^{\nu}-dy^{2}\,,
\ee
with $\epsilon$ being a very small real parameter and $ h_{\mu\nu}=h_{\mu\nu}(x^{\mu},y) $. We use transverse traceless gauge, $h_{\mu\nu}\rightarrow \bar{h}_{\mu\nu}$, to show that the metric fluctuation decouples from the fields~\cite{De}, leading to
\be
(\partial_{y}^{2}+4A^{'}\partial_{y}-e^{-2A}\square)\bar{h}_{\mu\nu}=0\,,
\ee
where $ \square=\eta^{\mu\nu}\partial_{\mu}\partial_{\nu} $. Introducing the $ z $ coordinate to make the  metric conformally flat with the choice  $ dz=e^{-A(y)}dy $ and making $ \bar{h}_{\mu\nu}=e^{ikx}e^{3A(z)/2}H_{\mu\nu}(z)  $, the above equation becomes
\be \label{eq31.2}
\left(-\dfrac{d^{2}}{dz^{2}}+U(z)\right)H_{\mu\nu}(z)=m^{2}H_{\mu\nu}(z)\,,
\ee
where $ U $ represents the stability potential, which can be written as
\be \label{eq31.3}
U(z)=\frac{9}{4}A_{z}^{2}+\frac{3}{2}A_{zz},
\ee
where $ A_{z} $ corresponds to the derivative with respect to the $z$ variable. As is known, the stability equation~\eqref{eq31.2} can be written in the form
\be \label{eq31.4}
\left(-\dfrac{d}{dz}-\frac{3}{2}A_{z}\right)\left(\dfrac{d}{dz}-\frac{3}{2}A_{z}\right)H_{\mu\nu}(z)=m^{2}H_{\mu\nu}(z)\,.
\ee
This factorization shows that the stability equation cannot support states with negative eigenvalues, indicating that the brane is stable under small fluctuations in the metric.

\begin{figure}[t]
\centering
\begin{subfigure}
\centering
{\includegraphics[width=6.5cm,height=4.5cm]{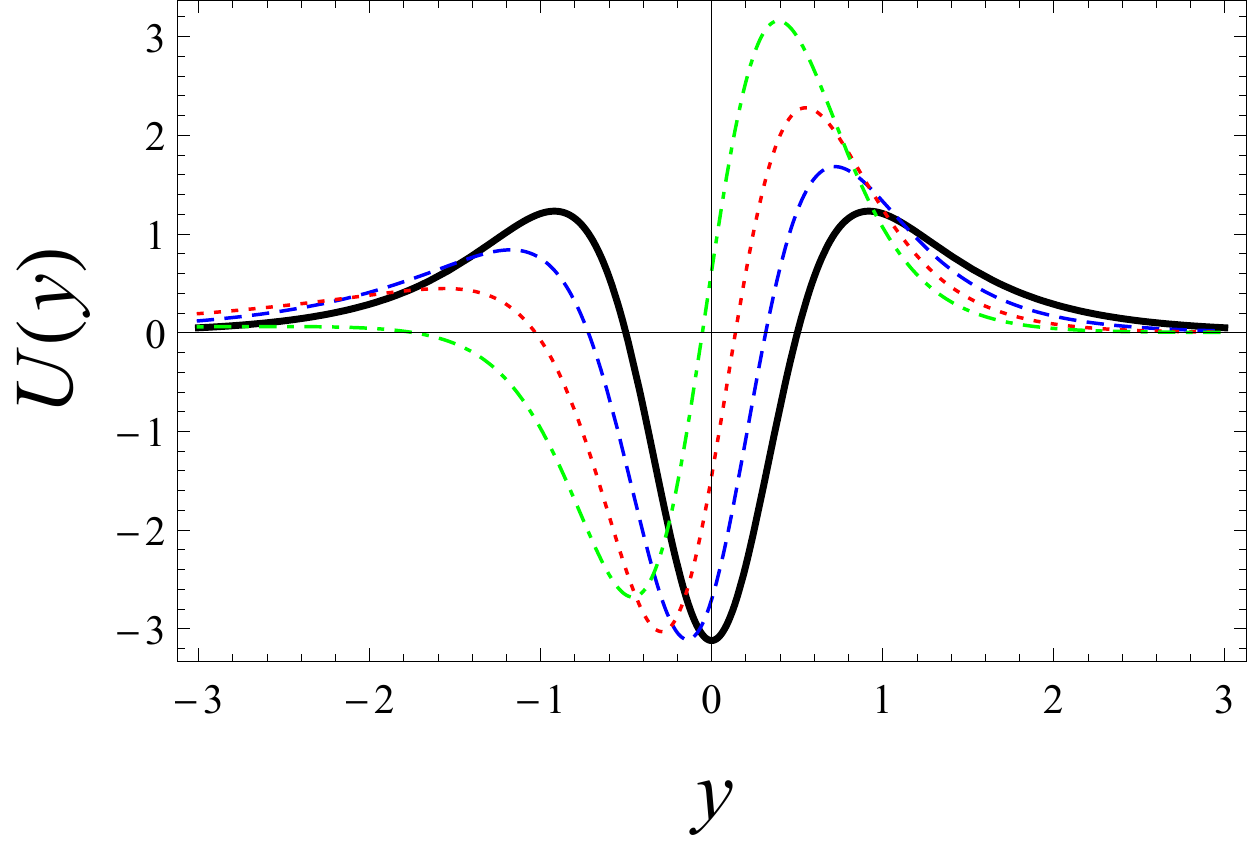}}
\end{subfigure}
\qquad
\begin{subfigure}
{\includegraphics[width=6.5cm,height=4.5cm]{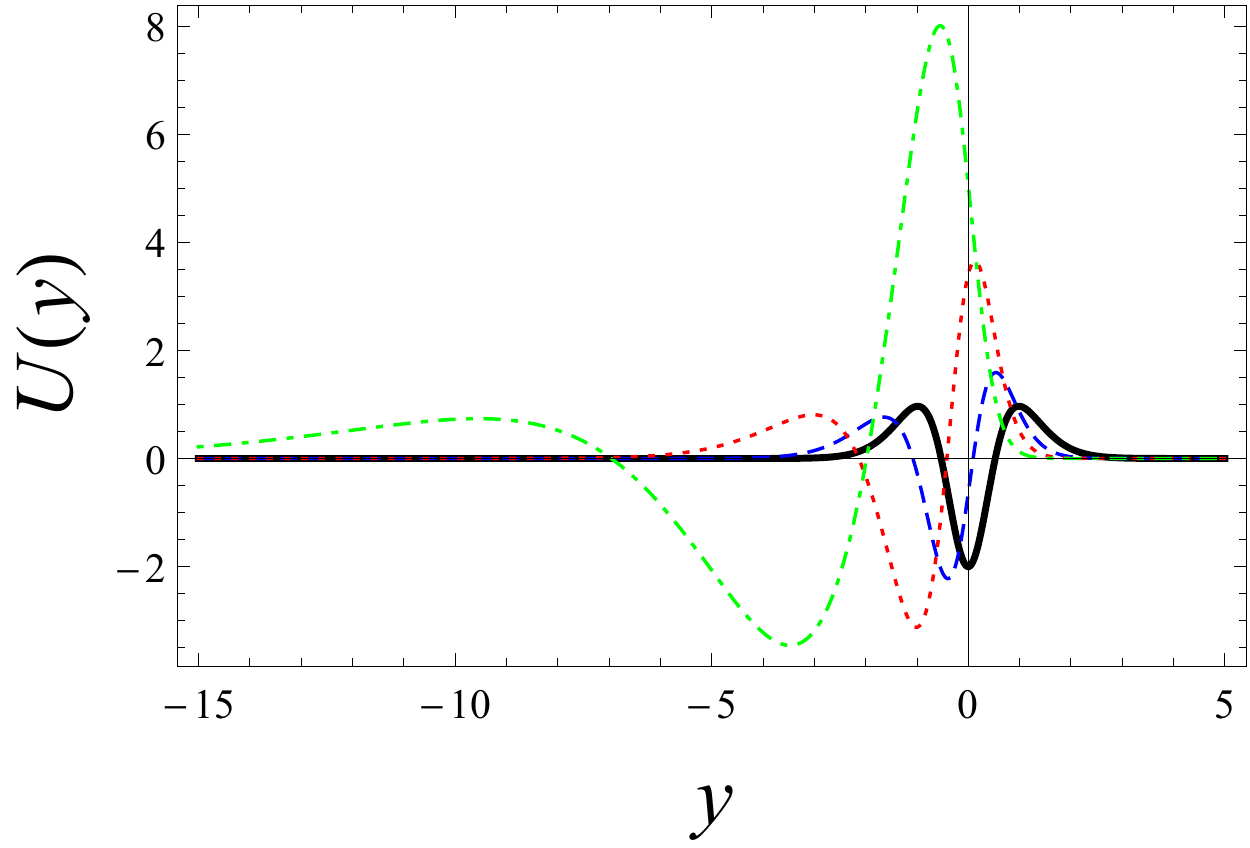}}
\end{subfigure}
\centering
\begin{subfigure}
{\includegraphics[width=6.5cm,height=4.5cm]{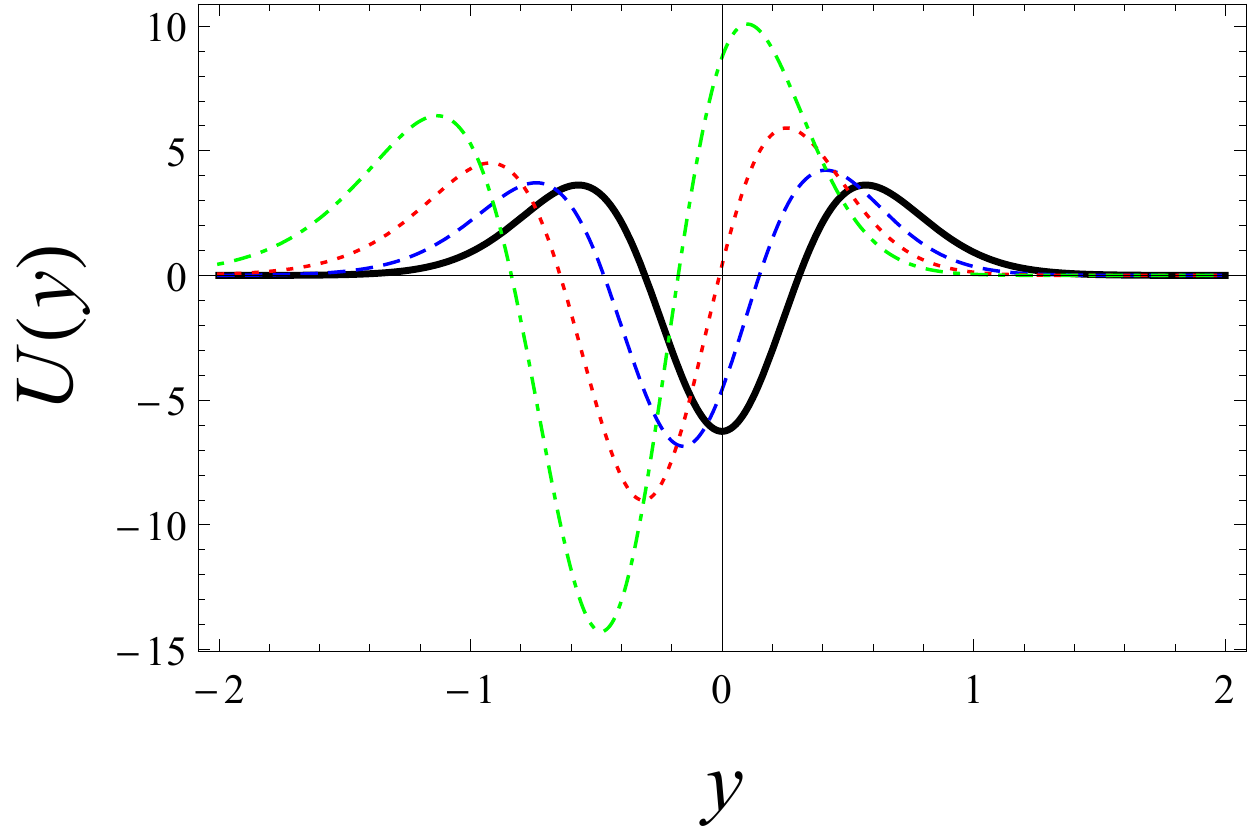}}
\end{subfigure}
\qquad
\begin{subfigure}
{\includegraphics[width=6.5cm,height=4.5cm]{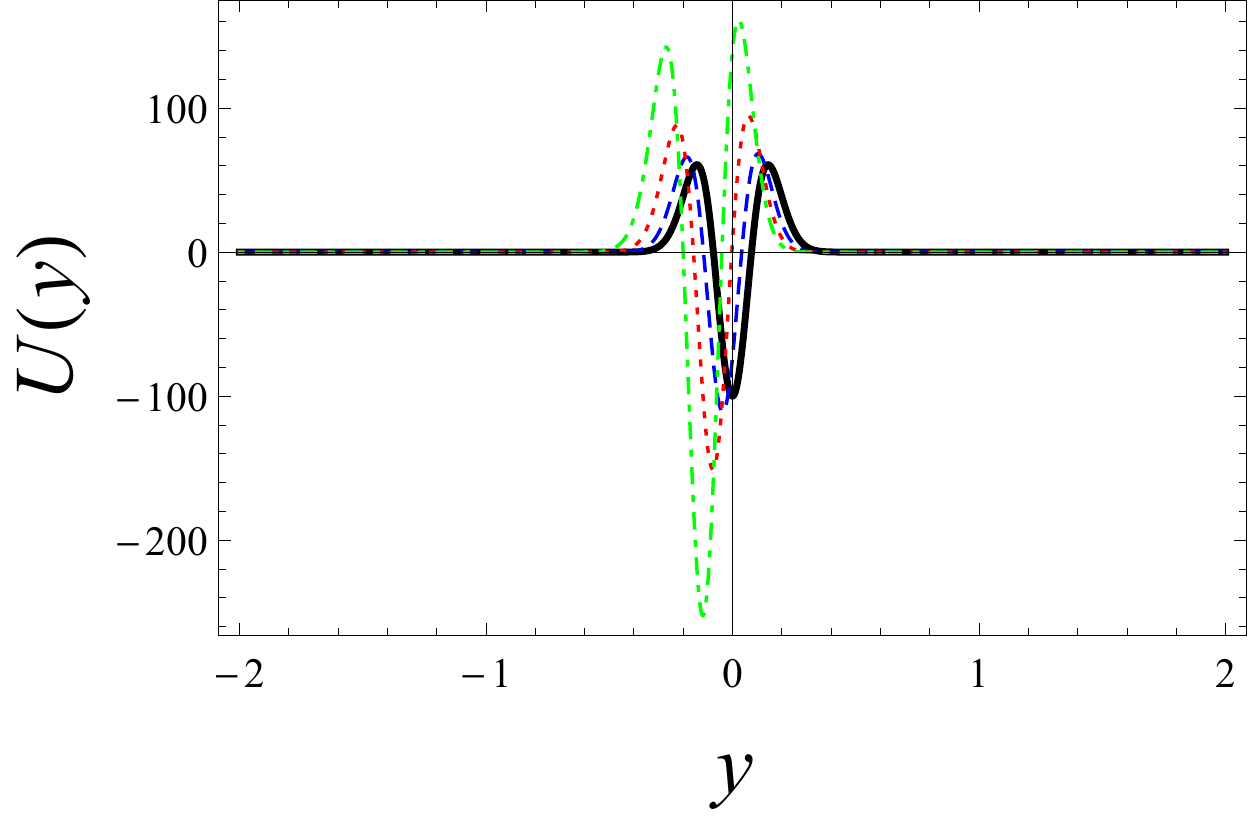}}
\end{subfigure}
\caption{Stability potential of the models.  At the top we show the model 1 (left) using the same conventions as Fig. \ref{Fig1} and the model 2 (right) using the same conventions as Fig. \ref{Fig2} with $ U (y)/2 $ in the last case. At the bottom we show the model 3 using the conventions of Fig. \ref{Fig3} (left) and Fig. \ref{figX} (right).}\label{Fig5}
\end{figure}
We have seen above that the study of stability of the metric under small perturbations leads to a Schr\"odinger-like equation, so the analysis of stability is done by inspection of the stability potential. In particular the three stability potentials arising from the models presented in the previous Sec. \ref{sec3} are depicted in Fig. \ref{Fig5}. They inform us that the three scenarios are stable under small fluctuations in the metric. In addition, they support zero mode (graviton zero mode) and no other bound state, as expected. The third and fourth panels are depicted for $\lambda=0.6$ and $0.9$, respectively, and for several values of $k$, to show the asymmetry behavior due to modifications of $k$, and the shrinking effect due to the increase of $\lambda$.

It is worth mentioning also that these stability potentials go to zero at infinity, so the mass spectrum is continuous without a gap, and it starts at $m = 0$. Moreover, the wave functions of the corresponding massive modes must have a plane wave profile.

\section{Generalized model} \label{sec5}

Let us now investigate how the procedure developed up to here can also be implemented in models with nonstandard dynamics. We consider the case in which the Lagrangian \eqref{eq3.2} changes to $ {\cal L}={\cal L}(\phi_{i},X_{ij}) $, where 
\be \label{eq5.1.1}
X_{ij}=\frac{1}{2}g^{ab}\triangledown_{a}\phi_{i}\triangledown_{b}\phi_{j}\,.
\ee
Using the same arguments that fields and warp function only depend on the extra dimension, Einstein's equations are now given by
\be \label{eq5.1.2}
A''=\frac{4}{3}X_{ij}{\cal L}_{X_{ij}}\,,
\ee
\be \label{eq5.1.3}
A'^{2}=\frac{1}{3}({\cal L}-2X_{ij}{\cal L}_{X_{ij}})\,,
\ee
with $ {\cal L}_{X_{ij}}=\partial {\cal L}/ \partial {X_{ij}} $. We can also implement a first-order formalism by considering the equation \eqref{eq3.6}, so the above equations become
\be \label{eq5.1.4}
\phi_{j}'{\cal L}_{X_{ij}}=W_{\phi_{i}}\,,
\ee
\be \label{eq5.1.5}
({\cal L}-2X_{ij}{\cal L}_{X_{ij}})=\frac{4}{3}W\,.
\ee

Let us now illustrate the procedure using a single field and considering that the Lagrangian is given by
\be \label{eq5.1.7}
{\cal L}(\phi,X)=-X^{2}-V(\phi)\,.
\ee
In this case, the Eq. \eqref{eq5.1.4} takes the form
\be \label{eq5.1.8}
\phi^{'}=W_{\phi}^{1/3}\,,
\ee
and the scalar potential can be written as
\be\label{ww}
V=\frac{3}{4}W_{\phi}^{4/3}-\frac{4}{3}W^{2}\,.
\ee
Now inspired by the Eq. \eqref{w} and the work \cite{Lo}, we choose the superpotential
\begin{equation}\label{w4}
W(\phi) =
\begin{cases} 
      -f(\phi)+\frac{3}{8}\sin^{-1}(\phi)+c\,, &  \phi^{2}\leq 1 \\
      f(\phi)+\frac{3}{8}\sgn(\phi)(\cosh^{-1}(|\phi|)+2\pi)+c\,, & \phi^{2}> 1
   \end{cases}
\end{equation}
where
\be
f(\phi)=-\frac{\phi}{8}\sqrt{|1-\phi^{2}|}(5-2\phi^2)\,,
\ee
and $ c $ is a real constant. Consequently, the Eq. \eqref{eq5.1.8} leads to the compact solution
\begin{equation}\label{s4}
\phi(y) =
\begin{cases} 
      \sin(y) & \text{for},\,\,  |y|\leq \frac{\pi}{2} \\
      \sgn(y) & \text{for}.\,\,  |y|> \frac{\pi}{2}
   \end{cases}
\end{equation} 
In this case the warp function is given by
\begin{equation}\label{A4}
A(y) =
\begin{cases} 
      -\frac{y^{2}}{8}-\frac{5}{24}\sin^{2}(y)+\frac{1}{24}\sin^{4}(y)-\frac{2cy}{3}\,, &  |y|\leq \frac{\pi}{2} \\
      -\frac{\phi}{8}|y|-\frac{16-3\pi^{2}}{96}-\frac{2cy}{3}\,. &  |y|> \frac{\pi}{2}
   \end{cases}
\end{equation}
By analyzing the minima of the scalar potential we have found a five-dimensional cosmological constant $ \Lambda_{5\pm}= -4/3 (c \pm 3 \pi /16)^{2} $. Thus, for $ 0 <\vert c \vert <3\pi/16 $ the model has different cosmological constants on each side of the brane connecting different $ AdS_{5} $ bulk spaces. 

\begin{figure}[!]
\centering
\begin{subfigure}
\centering
{\includegraphics[width=5cm,height=3.5cm]{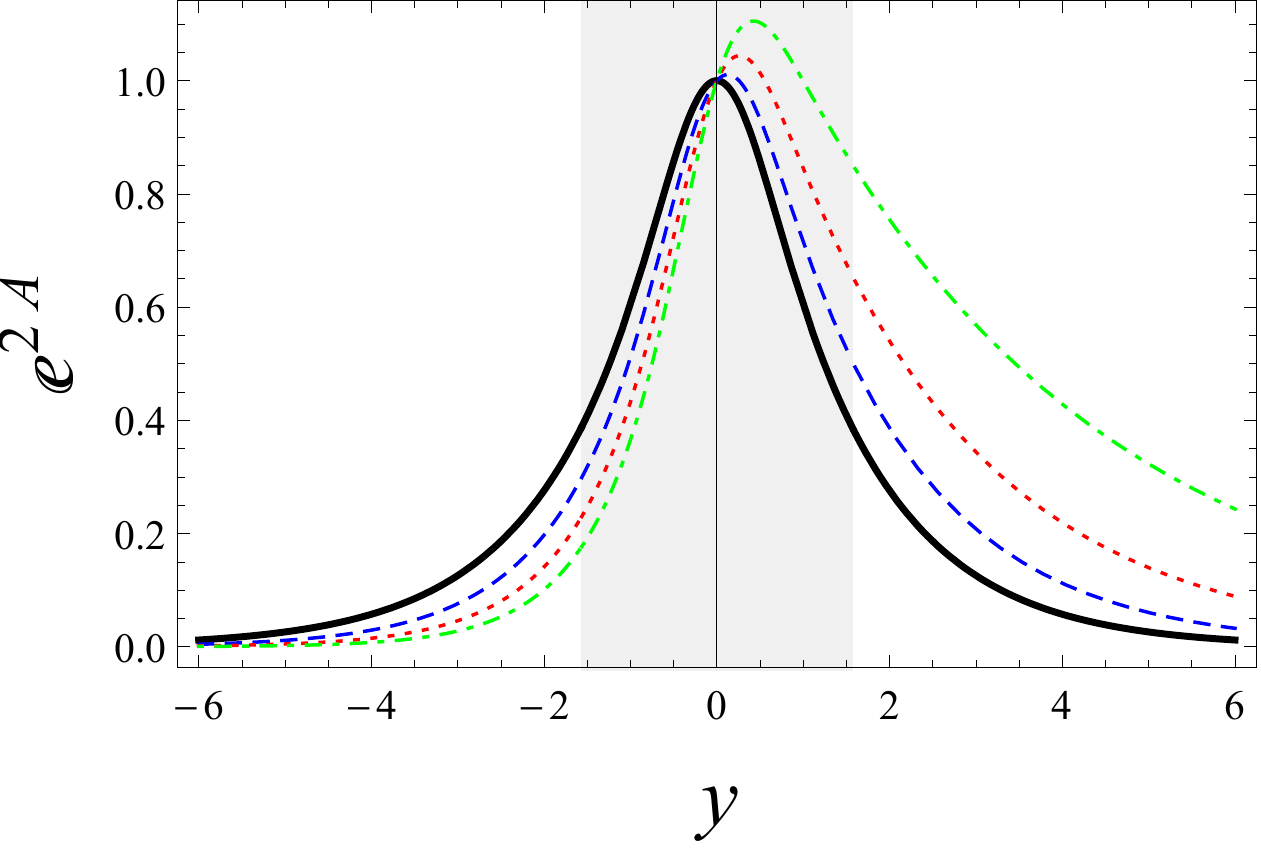}}
\end{subfigure}
\centering
\begin{subfigure}
{\includegraphics[width=5cm,height=3.5cm]{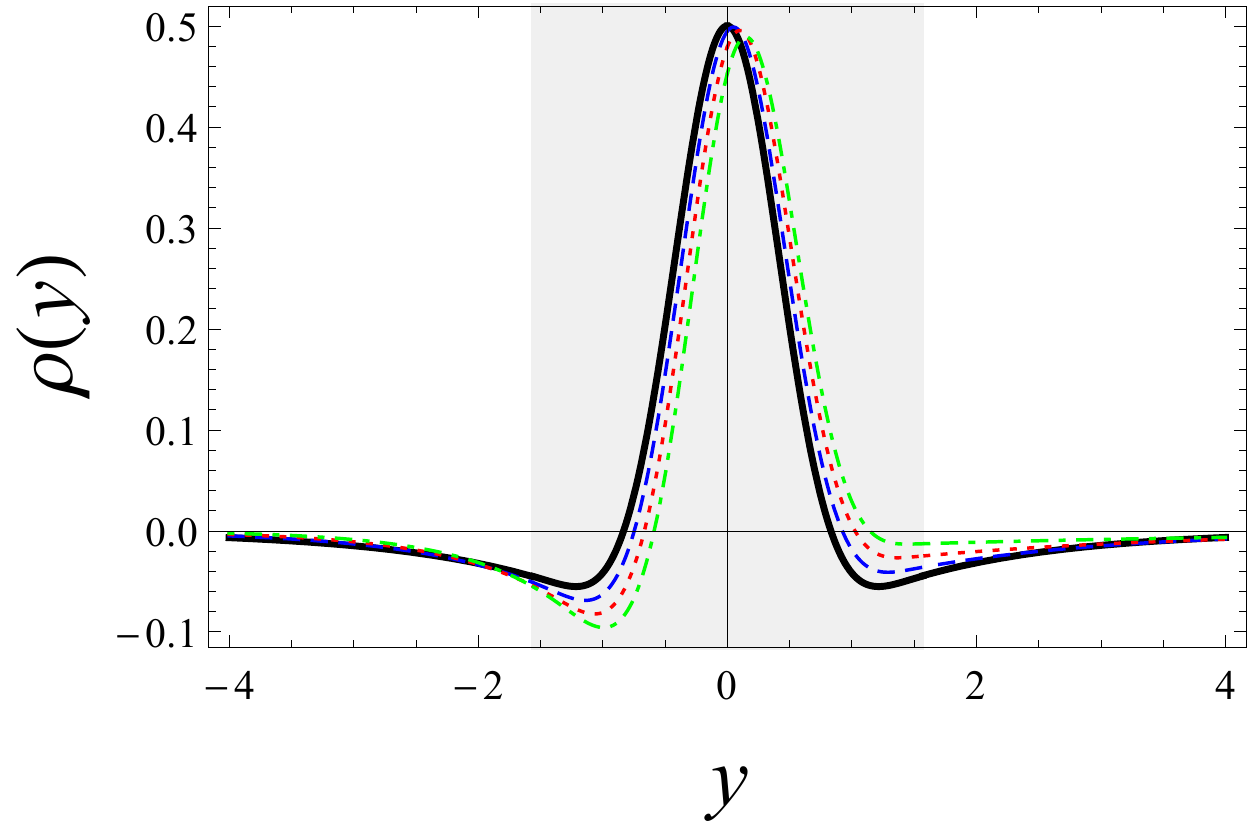}}
\end{subfigure}
\centering
\begin{subfigure}
{\includegraphics[width=5cm,height=3.5cm]{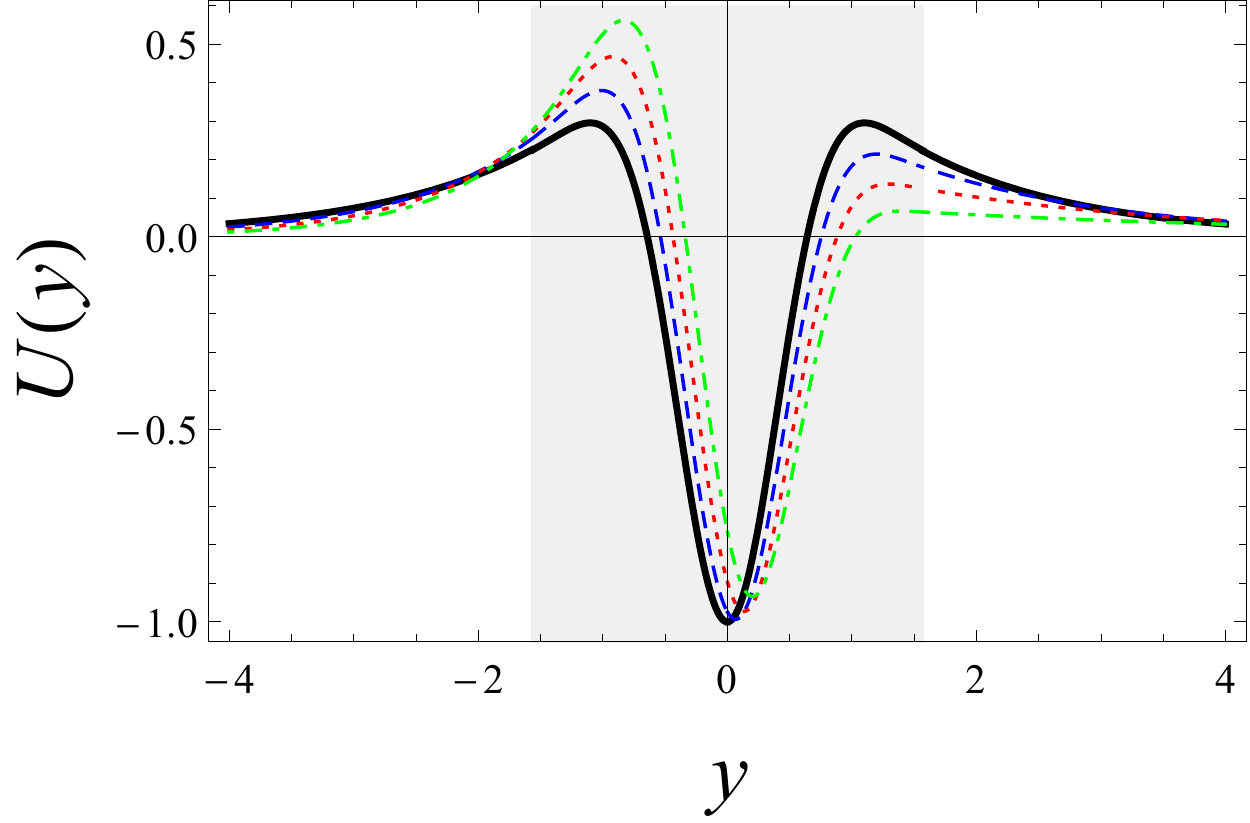}}
\end{subfigure}
\caption{The generalized model. The warp factor (left panel), energy density (middle panel) and stability potential (right panel) for $ c=0 $ (black, solid line), $ c=-\pi/25 $ (blue, dashed line), $ c=-2\pi/25 $ (red, dotted line) and $ c=-3\pi/25 $ (green, dot-dashed line).}\label{Fig7}
\end{figure}

Another important feature of brane is the energy density, it can be obtained from the energy-momentum tensor as
\be \label{eq5.1.6}
T^{00}=\rho=-e^{2A}{\cal L}\,.
\ee
Moreover, as shown in Ref.~{\cite{Lo}}, the stability study for $ N $ scalar fields in the generalized scenario does not contribute to destabilize the brane. Thus, stability potential in the nonstandard scenario is also given by Eq. \eqref{eq31.3}. In this sense, in Fig. \ref{Fig7} we depict the warp factor, energy density and stability potential for several values of $ c $, where the shaded portion indicates the region in which the scalar field is not constant. We see that in this generalized model the scalar field solution has the profile of a compact configuration, but this is not enough to make the brane compact. However, we notice that the parameter $c$ acts like the parameter $k$ that we introduced before in Sec. \ref{sec3} in the case of standard dynamics; in the generalized braneworld scenario, $c$ also contributes to make the brane asymmetric, so the idea of adding a constant to $W$ to make the brane asymmetric also works for the generalized model. 

We also observe from Eq. \eqref{ww}, that the presence of the second term in the definition of the potential $V$ will also lead us with nontrivially interacting models of $n$ scalar fields $\phi_1,\phi_2,...,\phi_n$ even in the case we take $W=W_1(\phi_1)+W_2(\phi_2)+\cdots+W_n(\phi_n)$, so the procedure explored in Sec. \ref{sec3} in the case of standard kinematics extends similarly to the above case of generalized models.

\section{Conclusion} \label{sec6}

In this work we have combined two distinct procedures to construct new models of asymmetric thick branes with $n$ scalar fields. We mixed and extended the investigations introduced in Refs.~{\cite{Ba3,Ahmed,Du3}} to construct new and stable braneworld scenarios that admit asymmetric thick branes, with the asymmetry controlled by a real parameter. For simplicity, we have considered the case of two scalar fields and illustrated our findings by examining several distinct models. We have also examined the stability of the models, verifying that they are stable under small fluctuations of the metric,  admitting only one bound state, the graviton zero mode. 

We extended our findings examining generalized models, in which the dynamics of the scalar field is of the nonstandard type. In this case, we have also shown that under specific circumstances, the scalar field potential can be described by a superpotential and the system admits first order differential equations that solve the equations of motion. We illustrated this possibility with another example, with the scalar field solutions being of the compact type. 

In the case of two scalar fields with standard kinematics, we have shown that the models are stable for several values of $k$, the parameter that controls the asymmetry of the brane. In the case of Model 3, in particular, besides $k$, we also have $\lambda$, the other parameter which controls the shrinking or stretching of the brane. We think it would be of interest to see how these parameters can be used to control cosmic acceleration as suggested before in Ref. \cite{Pa}, and in other circumstances of current interest in the braneworld scenario; see, e.g., Refs. \cite{CE2,C18}.

The present investigation has motivated us to investigate other related issues, and we are now studying the localization of four-dimensional gravity and the Newtonian limit for the case where the brane is embedded in a five-dimensional space {\cite{Fon}}; see also Ref. \cite{almeida}. Another possibility is to investigate the presence of fermions in these asymmetric thick brane models with several scalar fields, to see how the asymmetry and the addition of scalar fields can influence their entrapment. This can be implemented following the line of investigation of Refs.~{\cite{Mel,Li,Zhao}}. Another direction of current interest concerns the use of the procedure implemented of this work to describe aspects of mimetic gravity, both in the cosmology context \cite{mim1} and in the branework scenario \cite{mim2}. There are many other possibilities, and we think that the procedure presented in the current work will also work in other more general contexts, in particular in the case of metric-affine \cite{olmo}, Born-Infeld \cite{BI,BI2} and mimetic \cite{Mi} extensions that have been considered in the recent years.  These and other issues are currently under investigation and we hope to report on them in the near future.


\section*{Data availability}

The work contains analytical and numerical calculations.
The analytical calculations are fully explained in
the text of the manuscript. The numerical calculations are
also standard and consist in solving differential equations
numerically. The numerical and analytical calculations that
appear in the manuscript are all well established and can be
checked by others with no further difficulties.

\section*{Conflict of interest}
The authors declare that they have no conflict of interest.

\section*{Acknowledgments}
The authors would like to thank Laercio Losano, Matheus Marques and Roberto Menezes for discussions, and CAPES, CNPq and Paraiba State Research Foundation (Grant 0015/2019) for partial financial support. DB is also grateful to CNPq Grants 306614/2014-6 and 404913/2018-0.


\end{document}